\documentclass{aa}  

\usepackage{graphicx}
\usepackage{txfonts}
\newcommand{\bx}{\boldsymbol{x}}
\newcommand{\bv}{\boldsymbol{v}}

\newcommand{\Omegab}{\Omega_\mathrm{b}}

\newcommand{\Kpc}{~\mathrm{kpc}}
\newcommand{\pc}{~\mathrm{pc}}

\newcommand{\kmsec}{~\mathrm{km}~\mathrm{s}^{-1}}

\newcommand{\kmseckpc}{~\mathrm{km}~\mathrm{s}^{-1}~\mathrm{kpc}^{-1}}

\newcommand{\vc}{v_{\mathrm{c}}}

\newcommand{\Rg}{R_\mathrm{g}}

\newcommand{\RNum}[1]{\uppercase\expandafter{\romannumeral #1\relax}}

\newcommand{\degree}{^{\circ}}

\newcommand{\Js}{J_{\mathrm{s}}}

\newcommand{\Jf}{J_{\mathrm{f}}}
\newcommand{\Jp}{J_{\mathrm{p}}}

\newcommand{\ths}{\theta_{\mathrm{s}}}
\newcommand{\thf}{\theta_{\mathrm{f}}}
\newcommand{\thp}{\theta_{\mathrm{p}}}

\begin{document}

   \title{The signatures of the resonances of a \\large Galactic bar in local velocity space}

   \author{
    G. Monari \inst{1}
    \and
    B. Famaey \inst{2}
    \and
    A. Siebert \inst{2}
    \and 
    C. Wegg \inst{3}
    \and
    O. Gerhard \inst{4}}

    \institute{Leibniz Institut fuer Astrophysik Potsdam (AIP), An der Sterwarte 16, 14482 Potsdam, Germany \\
    \email{gmonari@aip.de}
    \and
    Universit\'e de Strasbourg, CNRS UMR 7550, Observatoire astronomique de Strasbourg, 11 rue de l'Universit\'e, 67000 Strasbourg, France
    \and
    Universit\'{e} C\^{o}te d'Azure, Observatoire de la C\^{o}te d'Azure, CNRS, Laboratoire Lagrange, Bd de l’Observatoire, CS 34229, 06304 Nice cedex 4, France
    \and
    Max-Planck-Institut f\"{u}r extraterrestrische Physik, Gießenbachstraße 1, 85748 Garching bei M\"{u}nchen, Germany
    }

   \date{Received xxxx; accepted xxxx}

 
  \abstract
   {The second data release of the Gaia mission has revealed a very rich structure in local velocity space. In terms of in-plane motions, this rich structure is also seen as multiple ridges in the actions of the axisymmetric background potential of the Galaxy. These ridges are probably related to a combination of effects from ongoing phase-mixing and resonances from the spiral arms and the bar. We have recently developed a method to capture the behaviour of the stellar phase-space distribution function at a resonance, by re-expressing it in terms of a new set of canonical actions and angles variables valid in the resonant region. Here, by properly treating the distribution function at resonances, and by using a realistic model for a slowly rotating large Galactic bar with pattern speed $\Omegab=39~\kmseckpc$, we show that no less than six ridges in local action space can be related to resonances with the bar. Two of these at low angular momentum correspond to the corotation resonance, and can be associated to the Hercules moving group in local velocity space. Another one at high angular momentum corresponds to the outer Lindblad resonance, and can tentatively be associated to the velocity structure seen as an arch at high azimuthal velocities in Gaia data. The other ridges are associated to the 3:1, 4:1 and 6:1 resonances. The latter can be associated to the so-called `horn' of the local velocity distribution. While it is clear that effects from spiral arms and incomplete phase-mixing related to external perturbations also play a role in shaping the complex kinematics revealed by Gaia data, the present work demonstrates that, contrary to common misconceptions, the bar alone can create multiple prominent ridges in velocity and action space.}


   \maketitle
%

\section{Introduction}

The local velocity distribution of stars near the Sun has long been known to exhibit clear substructures most likely caused by a combination of resonances with multiple non-axisymmetric patterns \citep[e.g.][]{Dehnen1998,Famaey2005} and of incomplete phase-mixing \citep{Minchev2009,Gomez2012} linked to external perturbations. The second data release from the Gaia mission \citep{Gaia} has now revealed this rich network of substructures with unprecedented details, displaying multiple clearly defined ridges in velocity space \citep{Katz2018,Ramos2018}, and even vertical velocity disturbances \citep{Antoja2018,Quillen2018a,Monari2018}, which have been associated to the perturbation of the disc by the Sagittarius dwarf galaxy \citep[e.g.][]{Laporte2018}, or the buckling of the Galactic bar \citep{Khoperskov2018}. As far as in-plane motions are concerned, it is also interesting to consider the distribution of stars in the space of actions, which are adiabatic invariant integrals of the motion constituting the natural coordinate system for Galactic dynamics and perturbation theory. \citet{Trick2018} produced such plots in various volumes around the Sun. For local stars ($d<200$~pc), they revealed several prominent ridges in the radial action distribution, among which a double-peak at the lowest border of the local azimuthal action (angular momentum) distribution, corresponding to the well-known Hercules moving group at low azimuthal velocities, and one at high angular momentum, corresponding to an arch at high velocities `covering' the velocity ellipsoid from above at $V \sim 40~\kmsec$, where $V$ is the heliocentric tangential velocity.

The Hercules moving group, in particular, has long been suspected to be associated to the perturbation of the potential by the central bar of the Galaxy \citep{Dehnen1999bar,Dehnen2000}. If the Sun is located just outside the bar's outer Lindblad resonance (OLR), where stars make two epicyclic oscillations while making one retrograde rotation in the frame of the bar, the Hercules moving group is naturally generated by the linear deformation of the unperturbed background phase-space distribution function (DF). This however implies a rather fast pattern speed for the bar, of the order of $\Omegab=55~\kmseckpc$ \citep[see also, e.g.,][]{Fux2001,Chakrabarty2007,Minchev2007,Minchev2010,Quillen2011,Antoja2014,Fragkoudi2019}, which past independent measurements also favoured \citep{Englmaier1999,Fux1999,Debattista2002,Bissantz2003}. By solving the linearized Boltzmann equation in the presence of the simple quadrupole bar potential of \citet{Dehnen2000}, we could show in \citet{Monari2016,Monari2017a,Monari2017} that the Hercules moving group was indeed naturally formed outside of the bar's OLR, and that its observed position in velocity space was varying as predicted by such a model. However, recent measurements of both the three-dimensional density of red clump giants \citep{Wegg2015} and the gas kinematics in the inner Galaxy \citep{Sormani2015} indicate that the pattern speed of the Galactic bar could be, in fact, significantly slower, as hinted by some older studies \citep{Weiner1999,Rodriguez2008,Long2013}. From dynamical modelling of the stellar kinematics in the inner Galaxy, \citet[][hereafter P17]{Portail2017} recently deduced a pattern speed of $\Omegab=39~\kmseckpc$. \citet{PerezVillegas2017} then showed, with orbit simulations in the potential of P17, that stars trapped at the co-rotation resonance of such a bar could also reproduce the Hercules position in local velocity space. Recently, using an update of the \citet{Tremaine1984} method -- similar to that of \citet{Debattista2002} -- on proper motion data from a combination of multi-epoch data from the VVV survey and Gaia DR2, \citet{Sanders2019} also found a pattern speed of $\Omegab=41 \pm 3~\kmseckpc$. Simultaneously, \citet{Clarke2019} also derived line-of-sight integrated and distance-resolved maps of mean proper motions and dispersions from the VVV Infrared Astrometric Catalogue combined with data from Gaia DR2, and found excellent agreement with a bar pattern speed of $\Omegab=37.5~\kmseckpc$. Studying in details all the dynamical effects of such a slowly rotating bar on the local velocity field is thus extremely timely. 

In \citet[][hereafter M17]{Monari2017b}, we developed an analytical method to capture the behaviour of the stellar phase-space DF at a resonance \citep[see also][]{Binney2018}, where the linearisation of the collisionless Boltzmann equation yields a divergent solution (problem of small divisors). This is indeed a fundamental difference between a model in which the Sun is located just outside of the bar's OLR and one where it is outside of the bar's corotation (CR) as in P17. Indeed, while the Hercules moving group is located outside of the trapping region in the former case, and can be treated through linearisation of the Boltzmann equation \citep{Monari2016,Monari2017a,Monari2017}, it is precisely located in the resonant trapping region for a slow bar like that of P17. However, with the M17 method, the deformation of velocity space induced by Dehnen's bar potential with a slow pattern speed was found to be rather minor. Here, by using the actual bar potential of P17 instead, and by computing the perturbed DF in the resonant regions, we confirm that the Hercules moving group can indeed be reproduced \citep{PerezVillegas2017}, and that many of the most prominent features in local action space can in principle be associated to the resonances of a slow bar model with $\Omegab=39~\kmseckpc$. 

In Sect.~2, we present the Galactic potential of P17 and our extraction of its Fourier modes. We then summarize in Sect.~3 the M17 method to treat the behaviour of the DF at resonances, and we apply in Sect.~4 this method to the main resonances of the $m=2$, $3$, $4$, $6$ Fourier modes of the P17 potential. We then validate our analytical treatment with numerical orbit integrations in Sect.~5, where we also compare the model to Gaia DR2 data. We conclude in Sect.~6.


\section{The bar potential}

\begin{figure}
\centering
\includegraphics[width=0.33\textwidth]{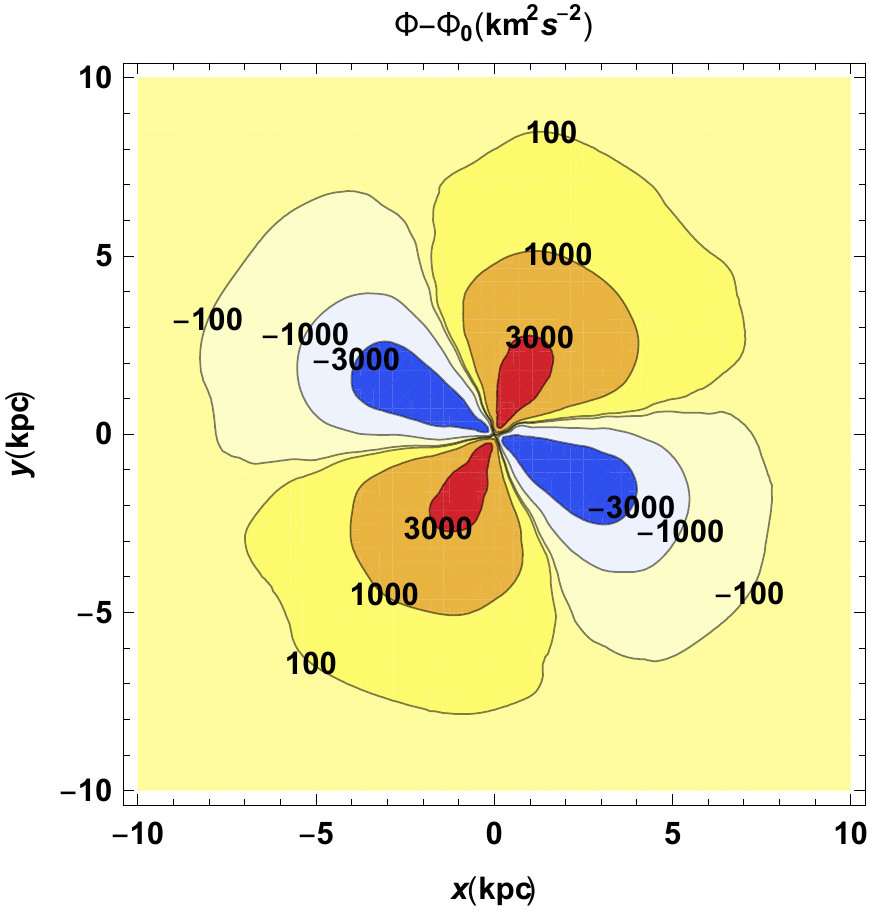}
\caption{The bar potential, i.e. the difference between the total P17 potential used in this work $\Phi$ and its axisymmetric $m=0$ part $\Phi_0$. The Sun in this model is placed at $(x,y)=(8.2~\Kpc,0)$. The bar is rotating clockwise. Its long axis (blue negative contours of $\Phi-\Phi_0$) is inclined at an angle of $28\degree$ from the line connecting the Sun to the center of the Galaxy.} 
\label{fig:potential}
\end{figure}

\begin{figure}
\centering
\includegraphics[width=0.45\textwidth]{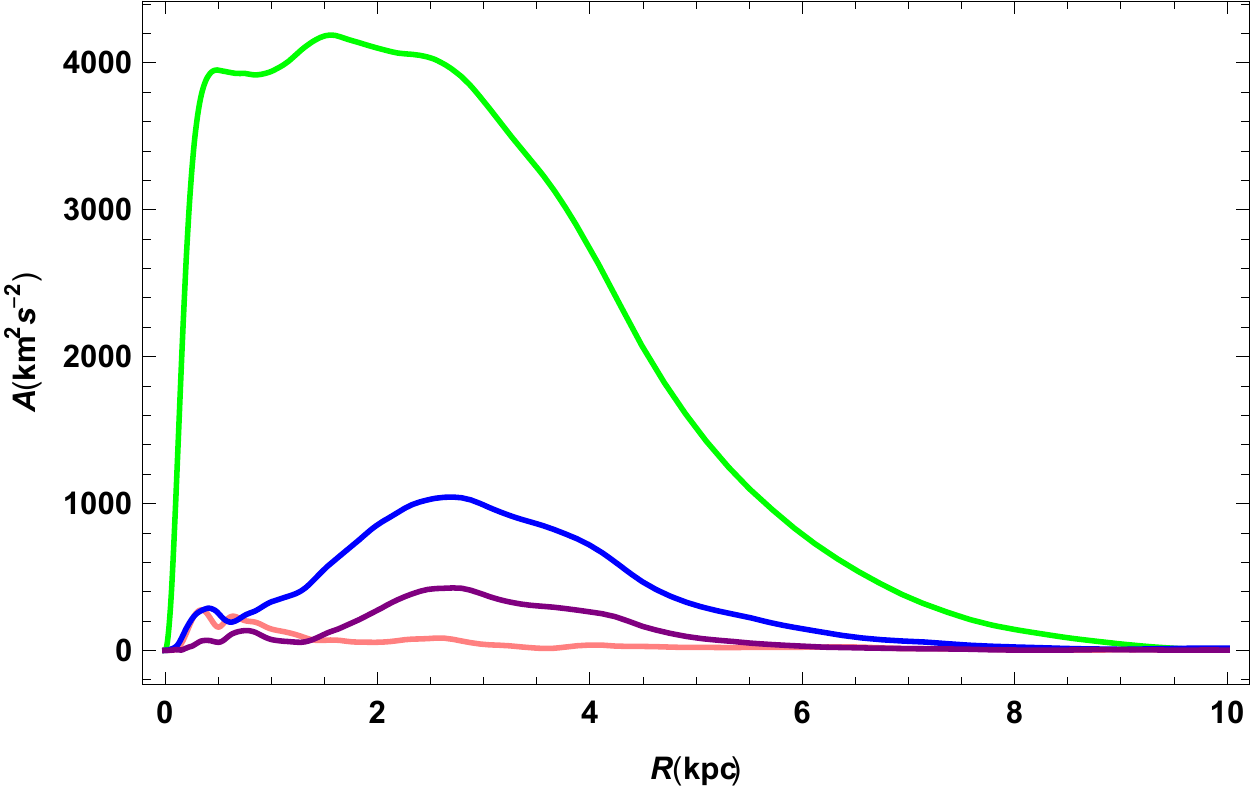}
\caption{Amplitude of the Fourier modes of the bar potential $A(R)$, as a function of the Galactocentric radius $R$. In green $m=2$, in blue $m=4$, in purple $m=6$, and in pink $m=3$.} 
\label{fig:potential_fourier}
\end{figure}

\begin{figure}
\centering
\includegraphics[width=0.33\textwidth]{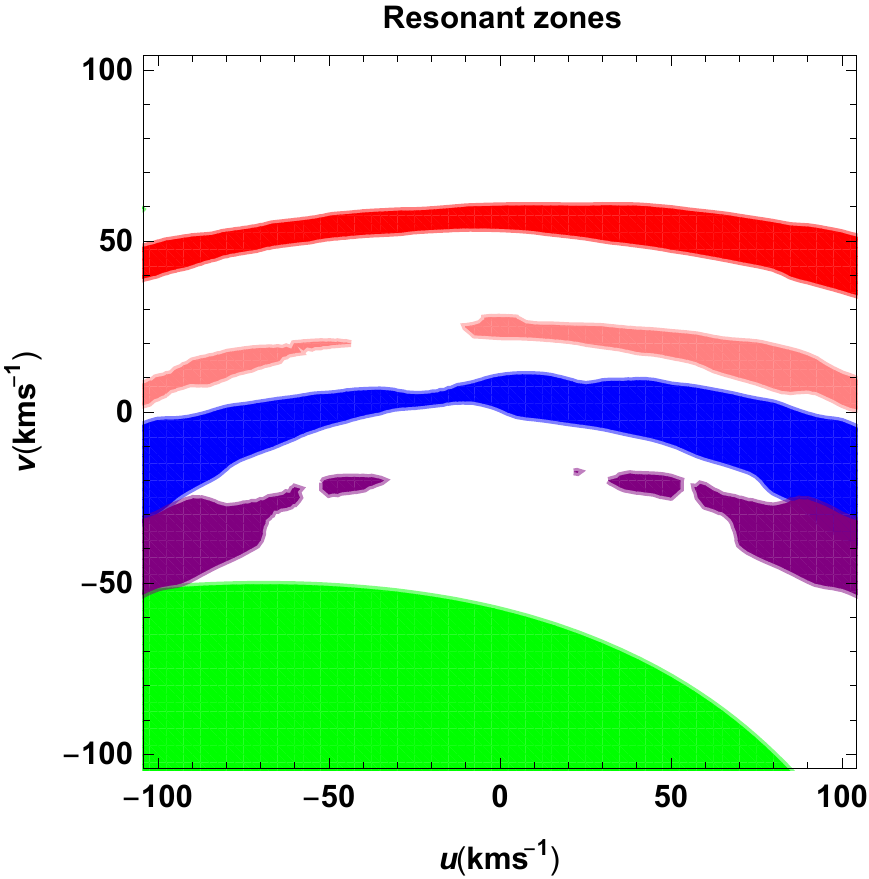}
\caption{Resonant zones in local velocity space (at the Sun's position) for different Fourier modes and resonances of the P17 bar, within the epicyclic approximation used in our analytical model (see Sects.~3 and 4). The resonant zones are defined as the regions where the energy parameter $k<1$ (see M17 and Sect.~3), i.e. the pendulum associated to each resonance is librating. In green and red, we display the corotation and 2:1 (OLR) resonance for the $m$=2 mode of the potential; in pink the 3:1 resonance of the $m$=3 mode; in blue the 4:1 resonance of the $m$=4 mode; in purple the 6:1 resonance of the $m$=6 mode. } 
\label{fig:resonant_zones}
\end{figure}

\begin{figure*}
\centering
\includegraphics[width=0.5\textwidth]{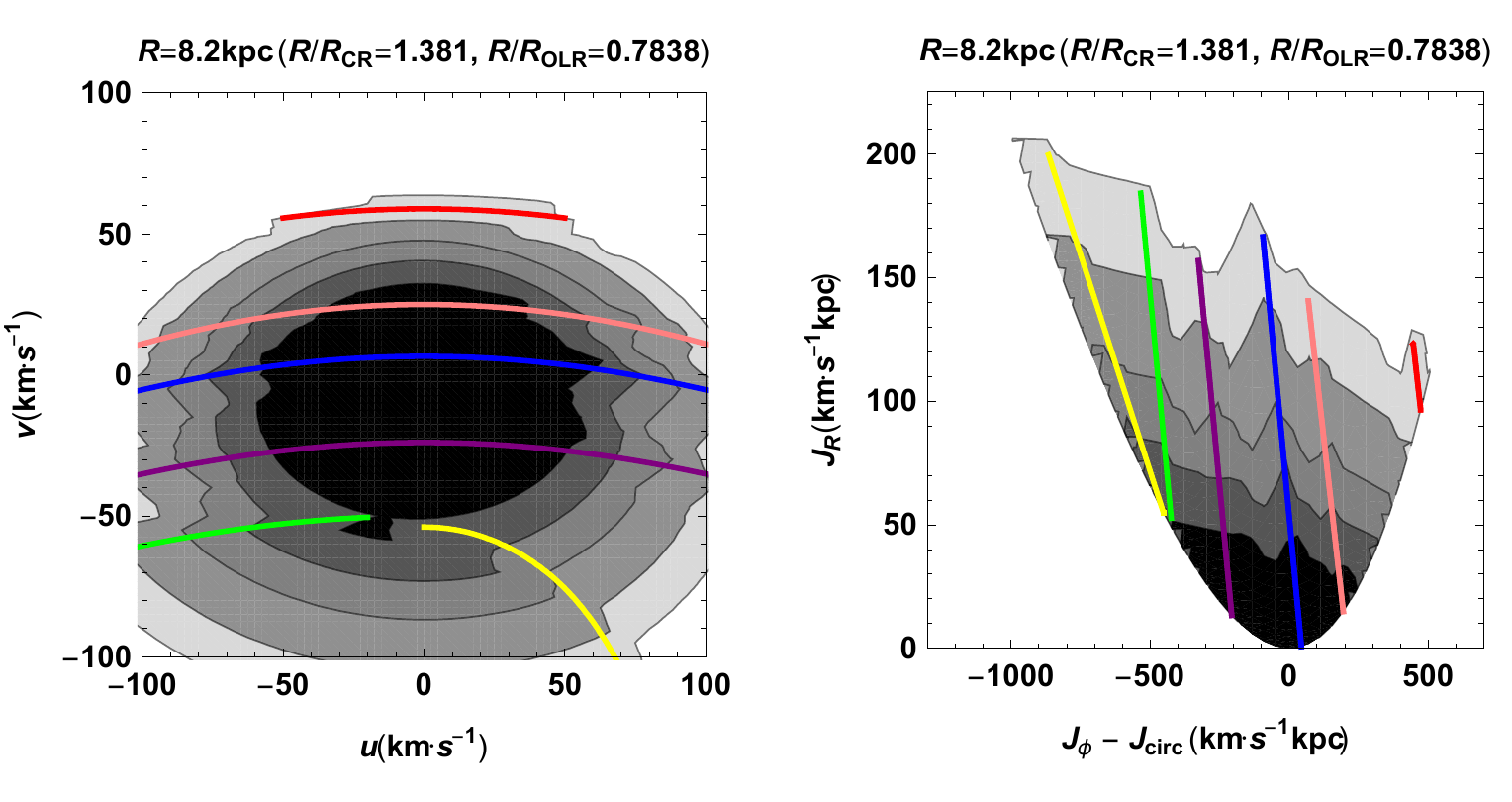}
\caption{Perturbed DF for stars located at the Sun's position in the model, obtained with the analytical method described in Sect.~\ref{Sect.:AA}. Left panel: velocity space $(u,v)$. Right panel: perturbed DF in axisymmetric epicyclic action space $(J_\phi-J_\text{circ}, J_R)$, where $J_\text{circ}$ is the angular momentum of a circular orbit at $R=8.2~\Kpc$. The contours include (from the darkest to the lightest) 50, 68, 80, 90, and 95 per cent of the stars. The radial velocity dispersion of the unperturbed DF $f_0$ at the Sun is $\sigma_R=45~\kmsec$. The colored lines correspond to each other in the two panels, and are plotted to identify the same features in velocity and action spaces. The green and yellow lines correspond to the corotation, and the red line to the 2:1 (OLR) resonance for the $m$=2 mode; the pink line corresponds to the 3:1 resonance of the $m$=3 mode, the blue line to the 4:1 resonance of the $m$=4 mode, and the purple line to the 6:1 resonance of the $m$=6 mode.} 
\label{fig:analytical_color}
\end{figure*}

We make use of the 2D Galaxy and bar potential $\Phi$ corresponding to the fit obtained by P17 in a range of radii between $R=0$ and $R=12~\Kpc$. As is evident from Fig.~\ref{fig:potential}, this potential is very different from the pure $m=2$ quadrupole that had been assumed in M17. In this potential model, the Sun is located at $R=8.2~\Kpc$ and at an azimuth inclined of $28\degree$ with respect to the long axis of the bar. We measure the azimuth $\phi$ from the long axis of the bar, so that the Sun is at $\phi=-28\degree$, i.e. in the direction opposite to the rotation of the Galaxy (and of the bar). The bar rotates at $\Omegab=39~\kmseckpc$, and the CR is located at $R \sim 6$~kpc.

The potential $\Phi(R,\phi)$ is defined on a grid, and the potential at any point in the Galaxy is obtained by spline interpolation. We also extract the Fourier components of the potential and in particular we use the $m=0$ mode as the `axisymmetric background' potential, and the $m=2$, $3$, $4$, and $6$ modes in Sect.~\ref{Sect.:AA}. In Fig.~\ref{fig:potential_fourier}, we display these four Fourier modes. We note that the $m=3$ mode should be zero for a model truly in equilibrium in the corotating frame, which is not exactly the case of the P17 model. We refer hereafter to the amplitude of the $m$-th mode of the potential as $\Phi_m$. In this way the background potential becomes $\Phi_0(R)$.

Using the background potential $\Phi_0$, we can define the Galactic circular frequency $\Omega(R)$ and epicyclic frequency $\kappa(R)$ in the usual way \citep[see e.g.][]{BT2008}, and the circular velocity curve $\vc(R)=R\Omega(R)$. 

For $R \gtrsim 12~\Kpc$, we assume an axisymmetric continuation of the form 
\begin{equation}    
    \Phi(R,\phi)=\Phi_0(R_\Phi)+\vc(R_\Phi)^2\ln(R/R_\Phi), \quad {\rm for}\, R\geq R_\Phi,
\end{equation}
i.e. a continuation corresponding to a flat circular velocity curve, and setting all non-axisymmetric modes $\Phi_{m>0}=0$ in these regions. However, we do not set the boundary $R_\Phi$ precisely at $12~\Kpc$, but slightly more inside, i.e. $R_\Phi=10~\Kpc$, 
because in this way we find that the transition is smoother, and at this radius the main $m>0$ components have already negligible amplitude. 

\section{The distribution function in resonant regions: action-angle formalism}\label{Sect.:AA}

Let $(J_R,J_\phi)$ be the radial and azimuthal actions and $(\theta_R,\theta_\phi)$ the canonically conjugated radial and azimuthal angles, defined in the background axisymmetric potential $\Phi_0$. A star's actions and angles (from now on AA) are combinations of the star's positions and velocities, and they are particularly convenient phase-space coordinates for several reasons, listed in \cite{BT2008}. In particular, the equilibrium axisymmetric background DF can be written purely as a function of the actions from Jeans' theorem, and they are the most convenient coordinates for perturbation theory. In simple words, the actions identify a star's orbit in phase-space, whilst the angles denote the phase of the star on that particular orbit. The larger the radial action $J_R$ is, the more energetic its radial excursions are, and the more eccentric the orbit is. The azimuthal action $J_\phi$ represents the vertical component of the angular momentum $L_z$. Here, we approximate the true values of the AA using the epicyclic approximation \citep[][M17]{BT2008}. In further work, we will extend the present modelling to more realistic AA variables obtained through a combination of the Torus Machinery \citep[e.g.][]{BinneyMcMillan2016} -- to go from AA variables to positions and velocities --, and of the `St\"ackel fudge' \citep[e.g.][]{Binney2012,SandersBinney2016} -- for the reverse transformation. For this reason, the results obtained in this analytic approach are still not fully quantitative, and we will confirm them in Sect.~5 with backward orbit integrations not making use of the AA variables in computing the response of the DF to the bar perturbation. Our analytic approach however offers a way to understand the physical mechanisms at play in the backward simulations.

Using the AA and the Galactic frequencies $\Omega$ and $\kappa$, we can define an `unperturbed' DF for the Galactic disc, i.e. a DF that would not change in time, because of the Jeans theorem, if there was no bar perturbation, which we denote $f_0(J_R,J_\phi)$. As in M17, we choose the quasi-isothermal DF defined by \citet{BinneyMcM2011}, with a scale length of 2~kpc, a velocity dispersion scale-length of 10~kpc, and a local velocity dispersion $\sigma_R(R_0)=45~\kmsec$ (slightly hotter than in M17).   

The response of the unperturbed DF to the bar potential is strongest at the resonances, that happen at the locations of phase space where
\begin{equation}\label{eq:resonance}
    l\omega_R(J_R,J_\phi)+m\left[\omega_\phi(J_R,J_\phi)-\Omegab \right]=0.
\end{equation}
where $\omega_R=\dot{\theta}_R$ and $\omega_\phi=\dot{\theta}_\phi$ are the orbital frequencies, and simply become, in the epicyclic approximation, $\omega_R = \kappa(\Rg)$ and $\omega_\phi = \Omega(\Rg) + [\kappa(\Rg)/J_\phi]J_R$, where $R_g(J_\phi)$ is the guiding radius \citep[e.g.,][]{Dehnen1999}.

In Eq.~\eqref{eq:resonance}, we will consider the main resonances of the $m=2, 3, 4, 6$ Fourier modes of the potential. For the $m=2$ mode, we shall take care of the $(l,m)=(0,2)$ resonance, namely the CR, and the $(l,m)=(1,2)$, namely the OLR. For the other modes, we will not consider the CR in the analytic approach, to avoid overlap of resonances, but we shall treat the $(l,m)=(1,3), (1,4), (1,6)$ resonances of the $m=3, 4, 6$ Fourier modes of the potential.

To study the response of $f_0$ near a resonance, we have to define, in each of the five resonant zones considered here, a new set of AA variables. We have to go through a first canonical transformation of coordinates, from the old AA $(J_R,J_\phi,\theta_R,\theta_\phi)$ to new `fast' and `slow' AA $(\Jf,\Js,\thf,\ths)$. The canonical transformation is \citep[][M17]{Weinberg1994}:
\begin{equation}
    \begin{aligned}
  \ths & =l\theta_R+m\left(\theta_\phi-\Omegab t\right),\quad & J_\phi & =m \Js, \\
  \thf &= \theta_R, \quad & J_R &= l\Js+\Jf,
\end{aligned}
\end{equation}
where $(l,m)=[(0,2), (1,2), (1,3), (1,4), (1,6)]$ for the CR and 2:1 OLR of the $m=2$ mode, 3:1 resonance of the $m=3$ mode, 4:1 of the $m=4$ mode, and 6:1 of the $m=6$ mode respectively. The angle $\ths$ is called the `slow angle' because it evolves slowly near a resonance, as is evident from the definition of the frequencies and that of the resonance in Eq.~\eqref{eq:resonance}. On the other hand, $\thf$ changes more rapidly and is called the `fast angle'. Physically, the slow angle $\ths$ represents the azimuth of the apocentre of the orbit in the reference frame where the unperturbed orbit would be close to the resonance. Therefore, $\ths$ represents the angle of precession of the orbit. At this point, expanding the potential in AA coordinates, it is possible to show that, near the resonances, $\Jf$ is almost constant along a star's orbit, and that one can average the Hamiltonian over the fast angles. For each value of $\Jf$, the motion in $\ths$ then becomes approximately a pendulum one, with $\Js(t)$ the momentum of the pendulum. The steps to show this are explained in detail in M17. One can then define a pendulum energy parameter $k$ depending on the phase-space coordinates, that determines whether the pendulum is `librating' (i.e. the $\ths$ angle oscillates back and forth between a maximum and a minimum without ever covering the whole $[0,2\pi]$ range), or `circulating' (i.e. $\ths$ has a motion that covers the whole range of angles). Stars at a position of phase-space where the $\ths$ pendulum is librating are called `trapped to the resonances'. In Fig.~\ref{fig:resonant_zones}, we display in local velocity space, at the position of the Sun, the five zones of trapping associated to the five resonances which we consider in the model of P17.

For a pendulum, it is of course possible to make a new canonical transformation defining the actual AA of the pendulum itself: $(\thp,\Jp)$. The way these depend on the phase-space coordinates changes according to whether the orbit is trapped or not. The pendulum AA define also the motion in $\Js(\thp,\Jp)$ nearby the resonances. We can therefore rewrite the unperturbed DF as $f_0(\Jf,\Js(\thp,\Jp))$. 

In M17 \citep[see also][]{Binney2018}, the perturbed DF close to the resonances was then defined as the original DF phase-mixed over the angles $\thp$
\begin{equation}
    f=\langle \, f_0(\Jf,\Js(\thp,\Jp)) \, \rangle,
\label{eq:librating}    
\end{equation}
where the average is done over the angle $\thp$. Outside the zone of resonance, the DF is instead described as
\begin{equation}
    f= f_0(\Jf, \, \langle\Js(\thp,\Jp)\rangle \, ),
\label{eq:circulating}
\end{equation}
where, in this case, the average represents the average $\Js$ of the circulating motion. Note that these recipes to obtain the perturbed DF are, in principle, different for every resonance also outside of the regions of trapping, while we want here to describe simultaneously five resonances of the Galactic bar. However, the value obtained outside the zone of resonance for $f$ is very similar in the case of the five resonances, because the $\Js$ oscillations amplitudes decrease fast going away from the resonant zone. It is therefore sufficient to take, outside the resonant regions of the phase space, the mean of the five DFs obtained for the five resonances. 

\section{Analytical results}\label{sect:analytical_results}

\begin{figure*}
\centering
\includegraphics[width=\textwidth]{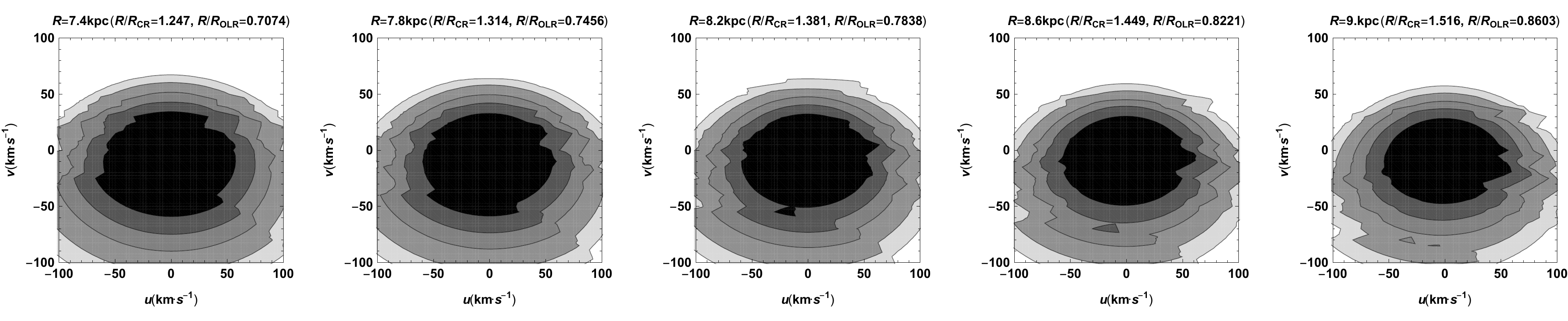}
\includegraphics[width=\textwidth]{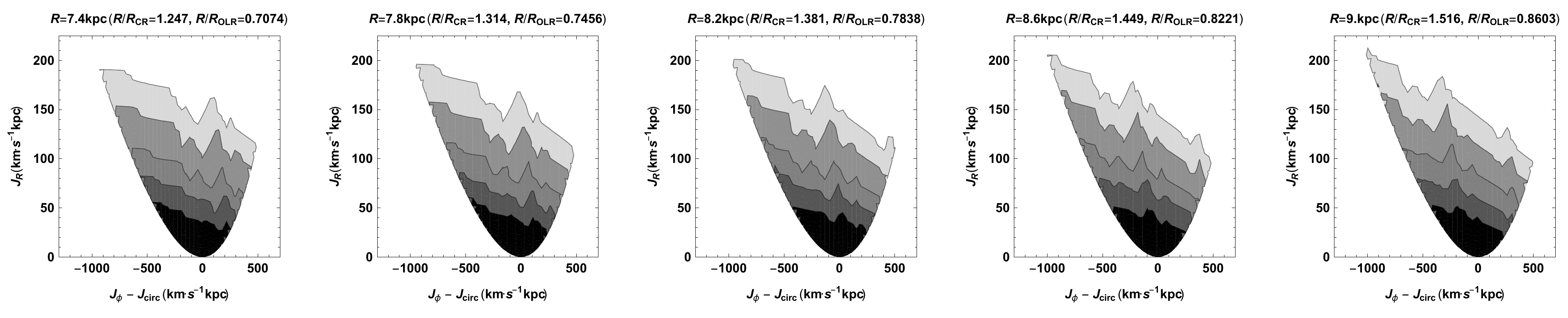}
\caption{As in Fig.~\ref{fig:analytical_color}, but for stars at different positions in the Galaxy (always aligned with the Sun and the Galactic center). Top row: DF in local velocity space. Bottom row: DF in local action space. The panels from left to right represent stars at $R=7.4$, $7.8$, $8.2$, $8.6$,  $9~\Kpc$.} 
\label{fig:analytical_radius}
\end{figure*}

\begin{figure}
\centering
\includegraphics[width=0.45\textwidth]{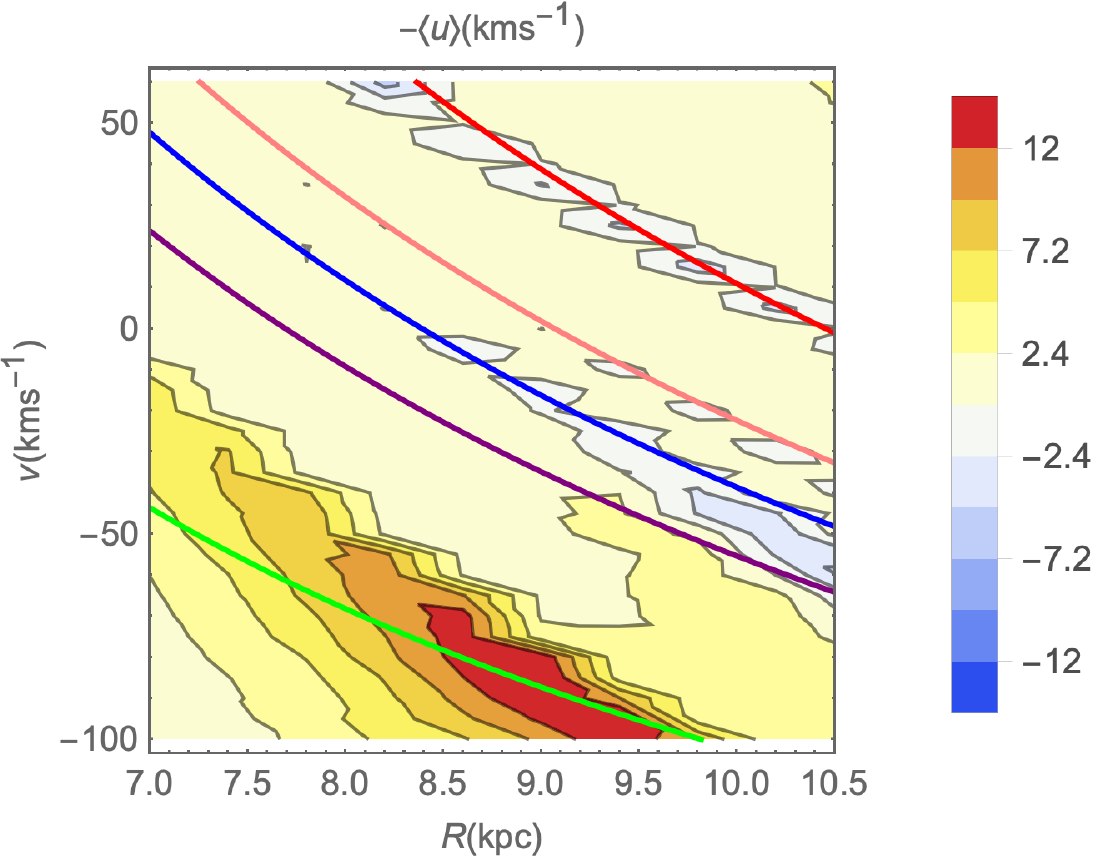}
\caption{Mean $-\langle u \rangle$ velocity (color bar) predicted by the analytical model in the $(R,v)$ space, for stars aligned with the Sun and the Galactic center (i.e. at the Solar azimuth). The bins are $\Delta R = 0.4~\Kpc$ and $\Delta v = 5~\kmsec$. The colored lines correspond to the positions in this space of the OLR (red line), 3:1 (pink line), 4:1 (blue line), 6:1 (purple line) and CR (green line). The velocity $-\langle u \rangle$ was chosen to facilitate the qualitative visual comparison with the figures in \cite{Laporte2018} and \cite{Fragkoudi2019}.}  
\label{fig:analytical_ridges}
\end{figure}

As in M17, we will concentrate hereafter on computing the perturbed DF at a few configuration space points, scanning through velocity space in order to define, at each phase-space point and for each resonance, the pendulum equation of motion for $\ths$ and its associated AA coordinates $(\Jp,\thp)$, allowing to compute the perturbed DF as described in the previous Sect. and as detailed in M17.

We start by plotting the perturbed DF as a function of the coordinates of velocity space at the position of the Sun in the model of P17. On the left panel of Fig.~\ref{fig:analytical_color}, we plot the DF in $(u,v)$ coordinates, namely the Galactocentric radial velocity in the direction of the Galactic center, and the peculiar tangential velocity with respect to the circular one at the Sun. 
As can clearly be seen on this figure, all five main resonances are present as deformations of the velocity distribution, that would be otherwise smooth. In particular the effect of the $m=2$ CR is to form an Hercules-like moving group \citep{PerezVillegas2017} with a peak at $(u,v)\simeq(-25,-50)~\kmsec$, and two asymmetric extensions at positive and negative $u$. At the same time, the 6:1 resonance forms a `horn'-like structure at larger $v$ and positive $u$, while the OLR forms a structure similar to the arch present at high azimuthal velocities in Gaia data (although not as strong as in the data).

To show even more clearly all these structures, on the right panel of Fig.~\ref{fig:analytical_color}, we show the DF in the $(J_R,J_\phi)$ action space of the unperturbed axisymmetric background. In this case, the deformation due to the CR appears clearly as two stripes at $J_\phi-J_\text{circ} \lesssim -500~\kmsec\Kpc$, corresponding to the deformation of the DF at positive and negative $u$ in the Hercules region of velocity space. In action space, the OLR is the pointy structure at $J_\phi-J_\text{circ} \simeq 500~\kmsec\Kpc$. In total, the bar alone thus creates no less than six prominent ridges in velocity and action space, at high as well as at low angular momenta, contrary to common misconceptions that the bar can only create one or two ridges, which is the case only when the bar is a pure $m=2$ mode \citep[see, e.g.,][]{Khanna}.

In Fig.~\ref{fig:analytical_radius}, we reproduce the same plots of the distribution in velocity and action spaces as a function of radius at the azimuth of the Sun. All structures shift with $R$ towards lower angular momenta as expected. In particular the effect of the CR is to form an Hercules-like moving group \citep{PerezVillegas2017} that shifts to lower $v$ and becomes less populated as $R$ increases, like the Hercules moving group does in the real data. 

It is also interesting to study how the different resonant ridges tend to move in terms of their radial velocity $u$ as a function of $R$. In \cite{Laporte2018} and \cite{Fragkoudi2019}, this has been illustrated from Gaia data as a mapping of the radial velocity as a function of position in the $R-v$ space (where $v$ refers here to the azimuthal velocity, even outside of the Solar neighbourhood). In Fig.~\ref{fig:analytical_ridges}, we produce a similar plot from our analytical model. Given the epicyclic approximation used in this approach, this should not serve as a direct quantitative comparison, but does capture the behaviour of the different resonances as a function of radius in the P17 model. In particular, one can note a splitting between the ridges associated to the 6:1 and CR resonances at $R \sim 9.5 \, {\rm kpc}$, reminiscent of Fig. 13 of \citet{Laporte2018}, as well as a ridge at high $v$ corresponding to the 2:1 OLR also present in the data. Perhaps more subtly, the ridges associated to the 3:1 and 4:1 resonances are very close to each other, and create a thick ridge at $R \gtrsim 9.5 \, {\rm kpc}$, which becomes thinner at smaller radii once the effect of the 3:1 resonance vanishes in this space, a feature which is also tentatively seen in the data. We note that, when the same P17 bar is given a pattern speed of $\Omegab=50~\kmseckpc$ (which is of course unphysical in the case of this particular bar model, because it would make the bar extend beyond its corotation radius), the ridge at high $v$ becomes inexistent, and the two ridges corresponding here to the CR and 6:1 resonances do not display their characteristic splitting around 9~kpc.

These plots, and in particular the one at $R=8.2$~kpc, thus allow us to physically understand the possible bar-related origin of many of the action space ridges identified by, e.g., \citet{Trick2018}. However, because of the epicyclic approximation used here, it will be useful to switch to orbit integrations in order to directly compare the effects of the bar to Gaia DR2 data.

\section{Backward integrations and comparison to Gaia}\label{Sect.:backwards}

In our analytical model hereabove, we approximated the values of the AA variables using the epicyclic approximation. One should however keep in mind that the position of the resonant zones in this analytic approach will depend on the ability of the AA epicyclic approximation to represent the true AA variables, which is the case only close to the center of the $(u,v)$-plane. While this drawback of the analytical method will be cured in further work, we chose to confirm our analytical results here with another technique to obtain the response of the DF to the bar perturbation, the `backward integrations' used for example by \cite{VauterinDejonghe1997} and \cite{Dehnen2000}. It was also recently used by \citet{Hunt2018} to explore the possibility that the Hercules moving group could be caused by the $m=4$ mode of a bar, with an amplitude similar to that found in $N$-body simulations. This method consists in integrating backwards the orbit of stars all at a certain point $\bx$ of configuration space now, but having different velocities on a grid in the $(u,v)$ plane. The method is based on the conservation of the phase-space density. Let us imagine that, at some time $t=t_1$ in the past, the amplitude of the bar was null, and then it started to grow with time until the current configuration at $t=0$. The value of the unperturbed DF $f_0$ at $t=t_1$ for the $i$-th velocity grid point (i.e. an orbit at $(\bx, \bv_i)$ at $t=0)$ was $f_0(\bx_{i,1},\bv_{i,1})$, where $\bx_{i,1}$ and $\bv_{i,1}$ are the initial position and velocity of the orbit. Then, because of the conservation of phase-space density, its current value at the point $\bx$ and the velocity $\bv_i$ is 
\begin{equation}
    f(\bx,\bv_i)=f_0(\bx_{i,1},\bv_{i,1}).
\end{equation}
This method has to assume an integration time and a law for the growth of the bar. In our case we assume the default values used by \cite{Dehnen2000}, i.e., 4 bar rotation for the integration time, a growth of the bar regulated by the polynomial law in Eq.~4 of \cite{Dehnen2000}, and a bar growth time of 2 bar rotations. We assume in this case the full numerical potential by P17, with its extension outside of $R=12~\Kpc$ as in the other Sections. Due to some heating induced by the bar growth, we choose a colder initial DF, with a local radial velocity dispersion of $35~\kmsec$, noting that the location of the structures in velocity space are independent of the velocity dispersion.

\begin{figure*}
\centering
\includegraphics[width=\textwidth]{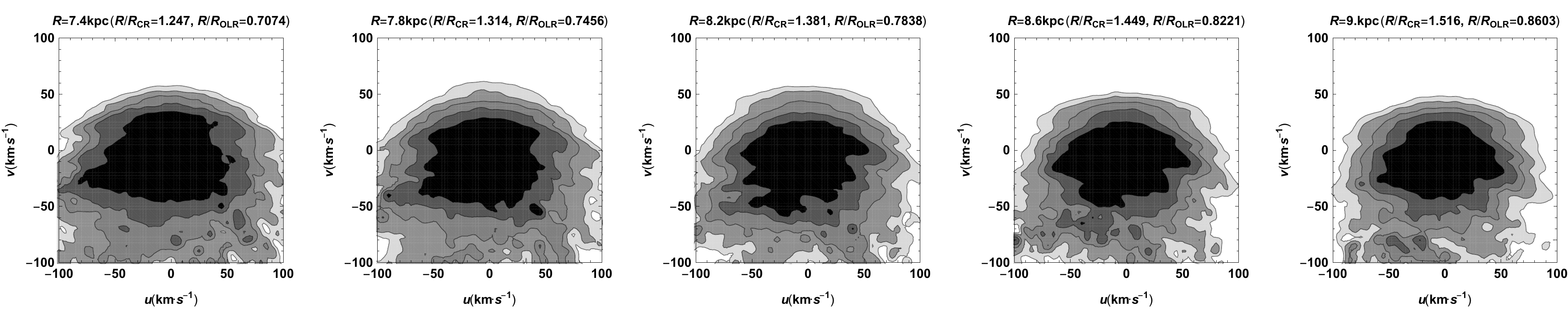}
\includegraphics[width=\textwidth]{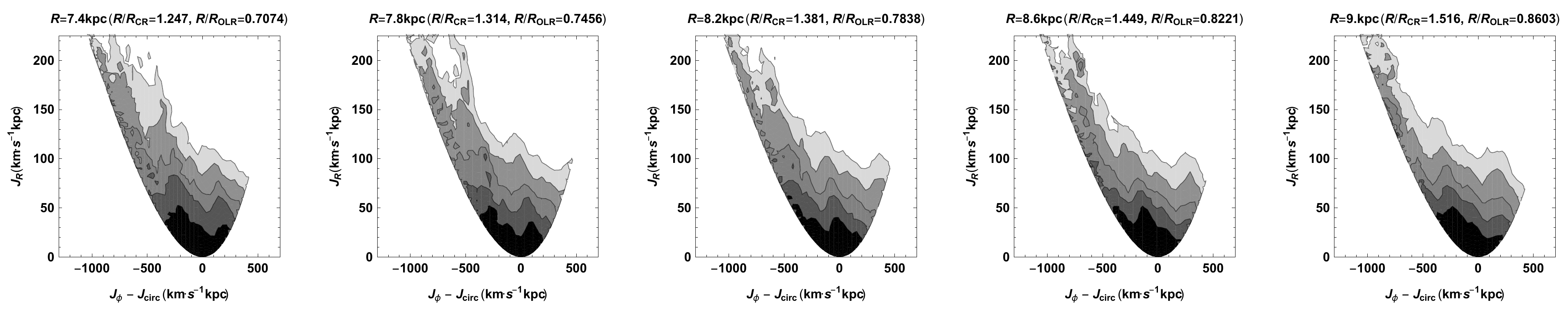}
\caption{As in Fig.~\ref{fig:analytical_radius}, but this time the DF is obtained using the backward integration technique described in Sect.~\ref{Sect.:backwards}. In this case, the initial value of the radial velocity dispersion of the unperturbed DF $f_0$ at the Sun was $\sigma_R = 35~\kmsec$.} 
\label{fig:back_radii}
\end{figure*}

The results of the backward integrations for the same points in configuration space as in Fig.~\ref{fig:analytical_radius} are presented in Fig.~\ref{fig:back_radii}. From this figure it is clear that the structures with this method are more complex for two main reasons: the ongoing phase-mixing, which adds noise to the backward integrations, especially for long integration times \citep{Fux2001}, and the effects of resonances not considered above, in particular the CR of the modes $m>2$, which tend to boost in action space the second ridge (green line on Fig.~4) associated to the Hercules stream. This Hercules group is centred at $(u,v)\simeq(-25,50)~\kmsec$ for $R=8.2~\Kpc$, and shifting towards lower (higher) $v$ as $R$ increases (decreases), as in the analytical prediction. In action space, one can also clearly identify the several ridges associated to the resonances described in Sect.~\ref{sect:analytical_results}, which are shifting in $J_\phi$ as a function of $R$.

\begin{figure*}
\centering
\includegraphics[width=0.6\textwidth]{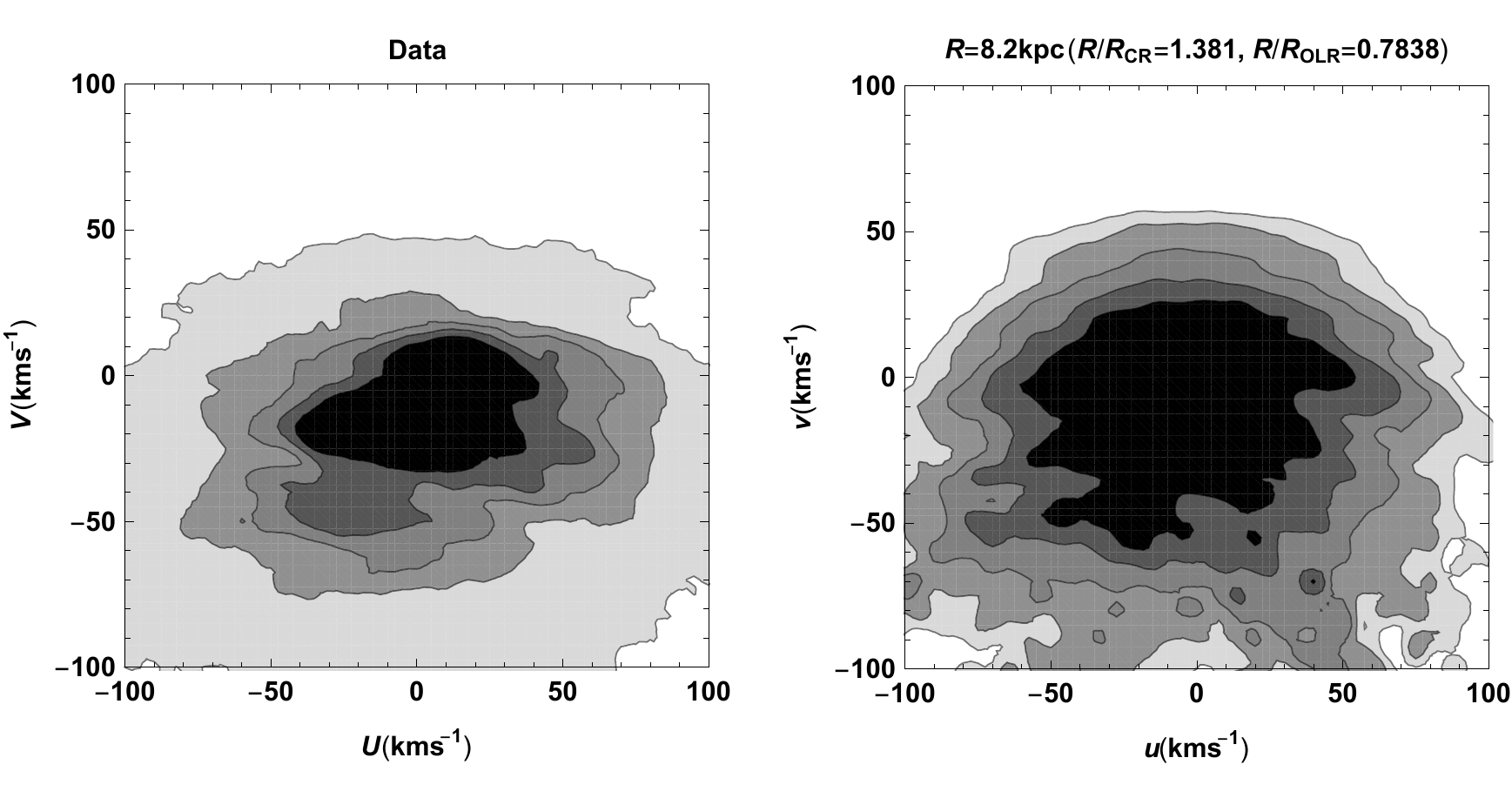}
\caption{Left panel: Observed local velocity $UV$ plane for Gaia DR2 stars inside a spherical volume of radius $0.2~\Kpc$, with the distances estimated by \citet{BailerJones2018}. The contours include (from the darkest to the lightest) 50, 68, 80, 90, and 99 per cent of the stars. Right panel: the $(u,v)$ velocity distribution at the Sun obtained with the backward integrations discussed in Sect.~\ref{Sect.:backwards} for the P17 model, i.e. the central panel (top row) of Fig.~\ref{fig:back_radii}. In this case, the contours include (from the darkest to the lightest) 50, 68, 80, 90, and 95 per cent of the stars.} 
\label{Gaia1}
\end{figure*}

\begin{figure*}
\centering
\includegraphics[width=0.6\textwidth]{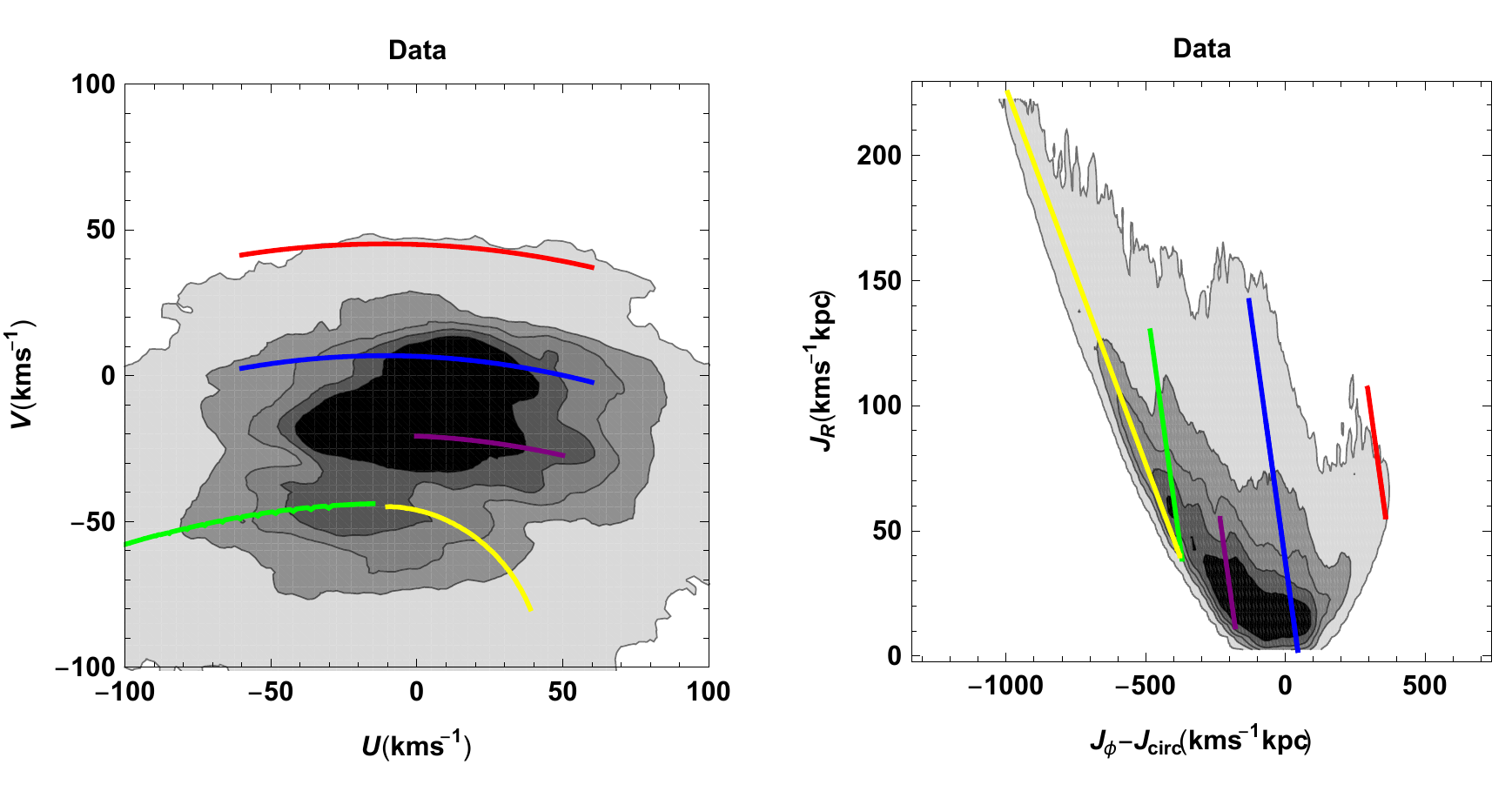}
\includegraphics[width=0.6\textwidth]{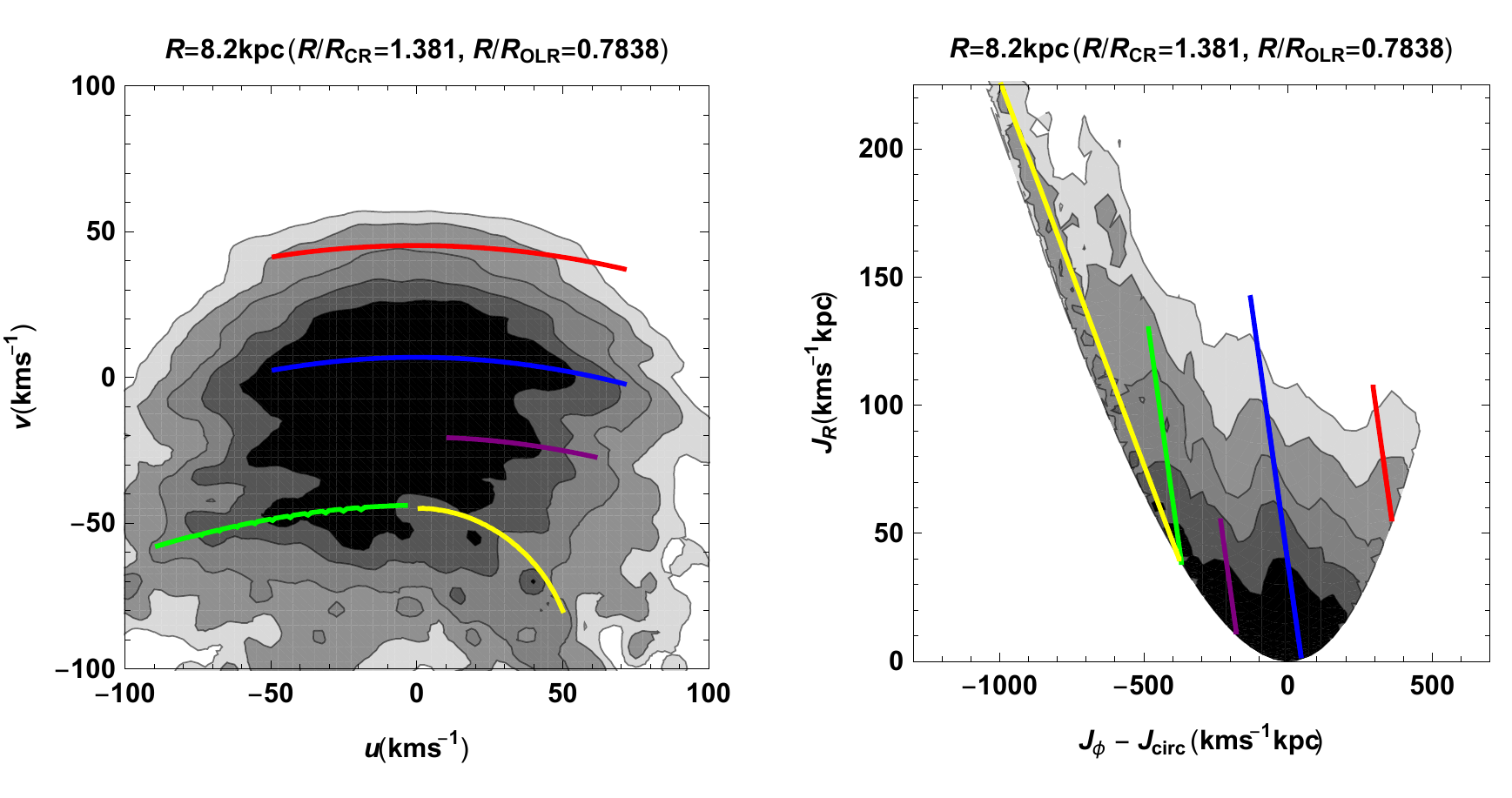}
\caption{Top row, left panel: Gaia $(U,V)$ kinematics in the solar neighbourhood from Fig.~\ref{Gaia1}; the contours contain 50, 68, 80, 90, and 99 per cent of the stars. Top row, right panel: as in the left panel, but this time the stars are plotted in the space of actions $(J_\phi-J_\text{circ},J_R)$; the actions are computed assuming the P17 background potential, the epicyclic approximation, and $(U_\odot,V_\odot)=(11.1,0)~\kmsec$; $J_\text{circ}$ is the angular momentum of a circular orbit at the Sun. Bottom row, left panel: $(u,v)$ distribution at the Sun obtained with backward integration from Fig.~\ref{Gaia1}; the contours contain 50, 68, 80, 90, and 95 per cent of the stars. Bottom row, right panel: as in the left panel, but this time the stars are plotted in the space of actions $(J_\phi-J_\text{circ},J_R)$. In all the plots the colored lines represent the same and corresponding positions in the $(u,v)$, $(J_\phi-J_\text{circ},J_R)$, and $(U,V)$ spaces. One can clearly identify the effect of the CR (Hercules, green line), 6:1 (`horn', purple line) and 4:1 (blue line) in both the model and data. The 2:1 resonance (red line) is also present in both, and most prominently seen in action space, but with a slight position mismatch between model and data, linked to the choice of $V_\odot$ and the shape of the circular velocity curve of the axisymmetric background (see text). Note also the prominent deformation in the velocity plane of the data at $(U,V) \sim (-35, -15) \kmsec$, namely the Hyades moving group, which is not reproduced by our bar-only model, and which has long been suspected to be related to a spiral perturbation.}
\label{Gaia}
\end{figure*}

In Fig.~\ref{Gaia1} we compare the actual velocity distribution of stars in the solar neighbourhood from the exquisite Gaia DR2 data \citep[][only stars with line-of-sight velocities in a spherical volume of 200~pc]{Gaia} with the one obtained from the backward integrations shown in the central panel, top row, of Fig.~\ref{fig:back_radii}. For the kinematics of the solar neighbourhood, we show the $(U,V)$ velocities, i.e. the Cartesian velocities in the rest frame of the Sun, with $U$ positive towards the Galactic center and $V$ positive towards the Galactic rotation. The $(u,v)$ velocities are related to the $(U,V)$ velocities, once the velocity of the Sun with respect to the local circular speed (or the `Local Standard of Rest') $(U_\odot,V_\odot)$ is known, which we discuss hereafter. The model's kinematics is hotter than that of the Gaia stars, and therefore for the latter we show the contour containing 99 per cent of the stars as the most external contour, instead of 95 per cent for the model. The comparison of the model with the data shows many similarities in the velocity distributions: the already cited Hercules moving group, the presence of the high $V$ arch (even more visible in action space in Fig.~\ref{Gaia}) which appears only in the 99 per cent contour of the Gaia stars at $V \sim 40~\kmsec$ (but is admittedly more marked in the data), two `horn'-like features, especially clear in the 68 per cent contours, at $(U,V) \sim (50,-25)~\kmsec$, and at $(U,V) \sim (40,-40)~\kmsec$. These could be connected to the structures in the 50 and 68 per cent contours in the models, at $(u,v) \sim (60,-30)~\kmsec$, and at $(u,v) \sim (50,-40)~\kmsec$. Finally, we can notice the presence of another large arc-like feature in the data, from $(U,V) \sim (-80,0)~\kmsec$ to $(U,V) \sim (60,0)~\kmsec$ and which could be identified with the arc-like feature in the model between $(u,v) \sim (-70,0)~\kmsec$ and $(u,v) \sim (70,0)~\kmsec$. 

We analyze in more detail these structures in Fig.~\ref{Gaia}, by associating them to features in epicyclic action space, using colored lines. We identify the high $V$ arch with a red line, which becomes particularly prominent in action space, the Hercules moving group with a green and a yellow line (for the negative and positive $u$ ridges respectively), the arch close to $V \sim 0$ with a blue line, and the most prominent horn at smaller $V$ with a purple line. 

To compare the colored lines from the $(U,V)$ plane of the data to the $(u,v)$ plane of the model, and convert both to action space, we have to assume a value for the Sun's peculiar motion $(U_\odot,V_\odot)$. In this case we use $U_\odot = 11~\kmsec$ \citep{Schonrich2010} and an unusually low value of $V_\odot=0$, which we will discuss below. As we have already seen in Sect.~\ref{sect:analytical_results}, the position of the structures in the $(u,v)$ plane in the bottom left panel correspond to ridges in action space, already identified in Sect.~\ref{sect:analytical_results}. In the top right panel we plot the action distribution for the data\footnote{The action distribution for the data is slightly more spread in $J_\phi$, even at small $J_R$, than in the model. This is due to the fact that, while in the model we plot the action space corresponding to stars all at one point in configuration space, the stars in the data are distributed in a sphere of finite volume and have observational errors.}, using the epicyclic approximation, the background potential of P17, and $V_\odot=0$. The distribution in this space presents ridges, already observed by other authors \citep{Trick2018}, and we see how nicely these correspond to structures identified in velocity space and present also in the model. Interestingly, we noted that the $m=3$ mode should be zero for a model truly in equilibrium in the corotating frame, and indeed this resonance is not standing out in the local data, even though it might have an effect at larger radii (see Fig.~6). On the other hand, the high angular momentum arch is prominent in action space both in the data and model, but its position is not perfectly reproduced. This is due to the choice of $V_\odot=0$. The latter is indeed the value that allows the best superposition with all the structures in the data and in the model, but for some individual structures such as this high $V$ arch, the superposition implies another value of $V_\odot$: the high $V$ arch is better fit for the more usual value $V_\odot=10~\kmsec$. Another possibility is that the P17 barred model is displaying deficiencies in explaining the data in the regions of velocity space where other effects such as phase-wrapping due to external perturbers \citep[e.g.][]{Laporte2018} become important.

The value $V_\odot = 0$ is of course unusual, and at odds with most estimates of the same parameter in the literature \citep[e.g.][]{DehnenBinney19982, Schonrich2010}. However, this unusual value could simply reflect the fact that the circular velocity curve of the Milky Way is somewhat different from that of the P17 model. Modifying only sightly the circular velocity curve $\vc(R)$ also modifies $\Omega(R)$ and $\kappa(R)$, and consequently the relative position of the resonances in local velocity space \citep[as well as the extension of the `gap' between the Hercules moving group and the main velocity ellipsoid, as was already noted by][]{Dehnen2000}. We performed a simple fit to see how we could modify $\vc(R)$ and $V_\odot$ to minimize the distance between the resonances and the position of the structures defined in the $(U,V)$ plane of the data. In this toy-model, we assumed for the background potential $\vc(R)=v_0(R/R_0)^\alpha$ where $R_0=8.2~\Kpc$ is the Sun's radius and $v_0=241 \kmsec$ is the value of $\vc(R_0)$ in the background model of P17. This simple fit shows that, for $V_\odot=8~\kmsec$ and a locally decreasing rotation curve ($\alpha=-0.1$), we get the best agreement between the position of the structures in the $(U,V)$ plane and the position of the resonances from the bar model of P17. However, all this could also depend on the precise value of the pattern speed, which might be slightly lower than assumed here \citep[e.g.,][]{Clarke2019}.

\begin{figure*}
\centering
\includegraphics[width=\textwidth]{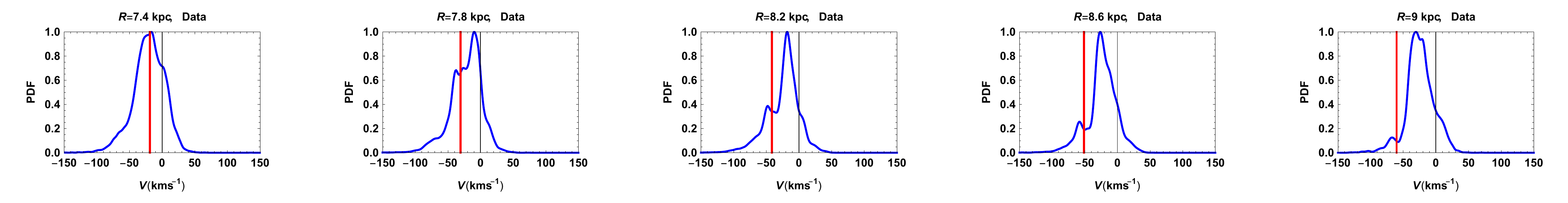}
\includegraphics[width=\textwidth]{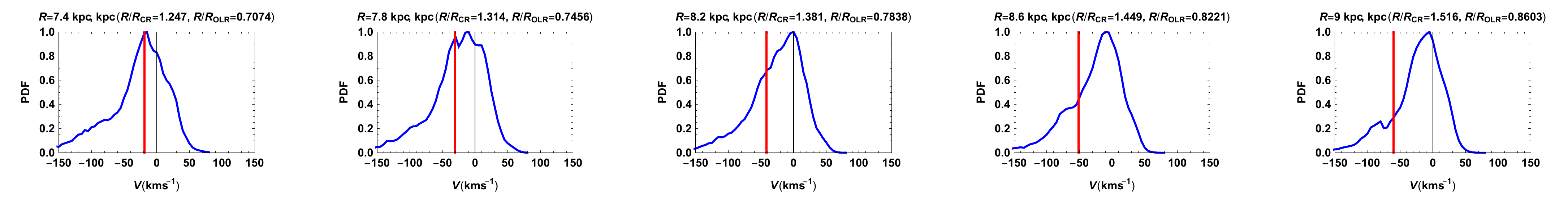}
\caption{Top row: Observed $V$ distribution for Gaia DR2 stars inside spherical volumes (of radius $0.2~\Kpc$) centred at different Galactocentric radii and along the line connecting the Galactic center to the Sun; the distances of the stars used here are those provided by \cite{BailerJones2018}. Bottom row: $v$ distribution obtained from the backward integration already shown in Fig.~\ref{fig:back_radii} and marginalizing along the $u$-axis (assuming $V_\odot=0)$. The red lines correspond to the same velocities on both rows, and to the same angular momentum as that of the gap at $R=8.2~\Kpc$.} 
\label{fig:v_radii}
\end{figure*}

Finally, in Fig.~\ref{fig:v_radii}, we present a further comparison of the model with the data, this time on a larger range of $R$, to show how the gap between the Hercules moving group and the main velocity mode changes with radius \citep[something that was already studied by][before the Gaia DR2]{Antoja2014,Monari2017,PerezVillegas2017}. We confirm the clear identification of the gap in the data and how it shifts with $R$ in the top row of Fig.~\ref{fig:v_radii}, following the vertical red line that we overplot on both the data and model. This line corresponds, at each $R$, to the velocity $v$ of stars that have the same angular momentum (i.e. the same guiding radius) as those that are at the gap velocity at $R=8.2~\Kpc$. In the bottom row, we show how the gap is also clear in the backward integration model, although less marked than in the data, in part due to a larger velocity dispersion, but also because of uncertainty on the unperturbed DF we assumed, and because the data must also be affected by other physical mechanisms. In particular, two bumps are especially clear in the data at 7.8~kpc, and not reproduced by the model. Again such limitations appear precisely in the regions of velocity space where other effects such as the inner resonances of outer spirals, or phase-wrapping due to external perturbers, become important.

\section{Discussion and conclusion}

In this work have used the analytical method recently developed in \citet[][M17]{Monari2017b} to study the response of the Galactic disc DF to the large bar Galactic potential model developed by \citet[][P17]{Portail2017}. This method is based on the use of action-angle (AA) coordinates and perturbation theory. 

Extracting from this Galactic potential model the Fourier modes of the bar, we have shown that in particular the $m$=2, $m$=4, and $m$=6 modes deform the disc DF in ways that resemble those shown by the second data release of the Gaia satellite around the Sun. The $m$=2 modes CR and OLR can be tentatively associated to features like the Hercules moving group (concave downwards in velocity space, as shown by the green and yellow lines on Figs.~4 and 9) and the high azimuthal velocity arch in the local velocity distribution of stars (albeit slightly less marked than in the data). Interestingly, the $6:1$ resonance of the  $m$=6 mode corresponds to the so called `horn' feature of local velocity space.

We also performed backward integrations using the whole Galactic bar model (no extraction of modes), showing that the same features obtained with the analytical method are also present in velocity and action space (but slightly offset because of the epicyclic approximation used for computing actions), further complicated by the presence of other modes of the bar and by ongoing phase-mixing.

The features obtained with the models can be seen most clearly in action-space, both in the model and in the Gaia data. The comparison between the model and the heliocentric velocity features in the data favours very small peculiar tangential velocities of the Sun. However, we showed that a small change of the circular velocity curve in the model can result in agreement with the data and a peculiar velocity of the Sun $V_\odot=8 \kmsec$ much more in line with other estimates in the literature. 

One caveat of using our method, as implemented in this work, is the use of the epicyclic approximation to estimate AA and orbital frequencies, which is valid only for orbits that deviate little from circularity. In the future we will update these results and extend them to the possibility to describe higher eccentricity orbits using more precise AA and frequencies estimation methods, like the St\"{a}ckel fudge \citep{SandersBinney2016} and the torus machinery \citep{BinneyMcMillan2016,Binney2018}. It would also be extremely important to extend perturbation theory methods, like the one used here -- or the linear theory to study the perturbations of the DF far from the resonances \citep{Monari2016}--, to the temporal evolution of the perturbations, and to be able to include non-equilibrium phase-mixing effects (Agobert et al. in prep.).

The purpose of this paper was to demonstrate which structures of (local and non-local) velocity space can be created by the resonances of a large Galactic bar {\it alone}. Future fits to larger volumes in configuration space, exploiting the full capabilities of the Gaia data, as well as future spectroscopic surveys such as WEAVE and 4MOST, will allow to test whether such structures are present in the data at all radii and azimuths, allowing to test such large bar models with a low pattern speed. However, one should keep in mind that other perturbations, mostly external perturbations and spiral arms, are present too, and that they cannot be ignored in a more thorough modelling. Importantly, their effects could explain some of the phase-space structures that the present bar-only model cannot. In particular, phase-wrapping due to external perturbations affect velocity space by creating ridges locally at $\sqrt{u^2+v^2} \gtrsim 40 \, \kmsec$ \citep{Minchev2009,Gomez2012,Antoja2018,Laporte2018}, while the bar model studied here does not create any structures beyond the region of velocity space delineated by its CR at low $v$ and OLR at high $v$. Some features identified in \citet{Ramos2018} and \citet{Laporte2018} are clearly outside of this zone. Moreover, \citet{Laporte2018} identified structures in {\it vertical} velocities in the $R-v$ space, which cannot be explained by the in-plane resonances of the bar studied here.
The same wave-like features have also been identified in the study of the mean vertical and radial velocities of the stars in Gaia DR1 and DR2 as a function of angular momentum (or guiding radius) by \cite{SchonrichDehnen2018} and \cite{FriskeSchonrich2019}. These wave-like features are partly aligned in mean radial and vertical velocity, and have also an azimuthal dependence compatible with $m=2$ and $m=4$ symmetries, and are reminiscent of the bending modes formed by satellite interaction on a galactic disc. This is also the favoured explanation that \cite{Carrillo2019} propose for many of the observed radial and vertical motions observed in Gaia DR2.
On the other hand, \citet{Quillen2018b} have shown that some ridges and arcs seen in the local velocity distribution could be caused by multiple spiral features with different pattern speeds. While our results remove, in principle, the need for an alternative explanation for some observed ridges that can be created by the bar resonances, it certainly does not exclude that other features are related to spiral arms. For instance, the Hercules moving group appears as a double-structure in the $UV$-plane of the data \citep[see also][]{Katz2018,Li2019}, and this splitting is not present in the $uv$-plane of the P17 bar-only model, thus asking for additional effects in order to split Hercules in two. One other obvious deficiency of the P17 bar model is its inability at explaining the strong Hyades overdensity at $(U,V) \sim (-35, -15) \kmsec$, which has long been suspected to be related to a spiral perturbation \citep[e.g.,][]{Quillen2005, Pompeia2011,McMillan2013}. The shape of the median velocity field as a function of position uncovered by \citet{Katz2018} is also highly suggestive of a spiral perturbation \citep[see also][]{Siebert2012}. Nevertheless, we conclude that the bar model studied here has some merits. Indeed, the most recent proper motion data within the inner Galaxy point towards a large bar with a low pattern speed \citep[e.g.,][]{Clarke2019,Sanders2019}. We showed here that the strength of the $m=$2, 4, and 6 modes of the P17 large bar model produces resonances with a noticeable amplitude \citep[see also][]{Hunt2018}, and that a low pattern speed puts those resonances at about the right positions in order to explain some of the most prominent ridges observed in local velocity and action spaces. Such a large bar model could thus serve as a basis model for future detailed study of other internal and external perturbations of the Galactic disc.

\begin{acknowledgements}
We thank the anonymous referee for a constructive report which greatly improved the article. BF and OG acknowledge hospitality at the KITP, supported by the National Science Foundation under Grant No. NSF PHY-1748958, in the final stages of this work. This work has made use of data from the European Space Agency (ESA) mission {\it Gaia} (\url{https://www.cosmos.esa.int/gaia}), processed by
the {\it Gaia} Data Processing and Analysis Consortium (DPAC,
\url{https://www.cosmos.esa.int/web/gaia/dpac/consortium}). Funding
for the DPAC has been provided by national institutions, in particular
the institutions participating in the {\it Gaia} Multilateral Agreement.
\end{acknowledgements}


\bibliographystyle{aa}
\bibliography{longbarbib}

\begin{thebibliography}{63}
\expandafter\ifx\csname natexlab\endcsname\relax\def\natexlab#1{#1}\fi

\bibitem[{{Antoja} {et~al.}(2014){Antoja}, {Helmi}, {Dehnen}, {Bienaym{\'e}},
  {Bland-Hawthorn}, {Famaey}, {Freeman}, {Gibson}, {Gilmore}, {Grebel},
  {Kordopatis}, {Kunder}, {Minchev}, {Munari}, {Navarro}, {Parker}, {Reid},
  {Seabroke}, {Siebert}, {Steinmetz}, {Watson}, {Wyse}, \&
  {Zwitter}}]{Antoja2014}
{Antoja}, T., {Helmi}, A., {Dehnen}, W., {et~al.} 2014, \aap, 563, A60

\bibitem[{{Antoja} {et~al.}(2018){Antoja}, {Helmi}, {Romero-G{\'o}mez}, {Katz},
  {Babusiaux}, {Drimmel}, {Evans}, {Figueras}, {Poggio}, {Reyl{\'e}}, {Robin},
  {Seabroke}, \& {Soubiran}}]{Antoja2018}
{Antoja}, T., {Helmi}, A., {Romero-G{\'o}mez}, M., {et~al.} 2018, \nat, 561,
  360

\bibitem[{{Bailer-Jones} {et~al.}(2018){Bailer-Jones}, {Rybizki}, {Fouesneau},
  {Mantelet}, \& {Andrae}}]{BailerJones2018}
{Bailer-Jones}, C.~A.~L., {Rybizki}, J., {Fouesneau}, M., {Mantelet}, G., \&
  {Andrae}, R. 2018, VizieR Online Data Catalog, 1347

\bibitem[{{Binney}(2012)}]{Binney2012}
{Binney}, J. 2012, \mnras, 426, 1324

\bibitem[{{Binney}(2018)}]{Binney2018}
{Binney}, J. 2018, \mnras, 474, 2706

\bibitem[{{Binney} \& {McMillan}(2011)}]{BinneyMcM2011}
{Binney}, J. \& {McMillan}, P. 2011, \mnras, 413, 1889

\bibitem[{{Binney} \& {McMillan}(2016)}]{BinneyMcMillan2016}
{Binney}, J. \& {McMillan}, P.~J. 2016, \mnras, 456, 1982

\bibitem[{{Binney} \& {Tremaine}(2008)}]{BT2008}
{Binney}, J. \& {Tremaine}, S. 2008, {Galactic Dynamics: Second Edition}
  (Princeton University Press)

\bibitem[{{Bissantz} {et~al.}(2003){Bissantz}, {Englmaier}, \&
  {Gerhard}}]{Bissantz2003}
{Bissantz}, N., {Englmaier}, P., \& {Gerhard}, O. 2003, \mnras, 340, 949

\bibitem[{{Carrillo} {et~al.}(2019){Carrillo}, {Minchev}, {Steinmetz},
  {Monari}, {Laporte}, {Anders}, {Queiroz}, {Chiappini}, {Khalatyan}, {Martig},
  {McMillan}, {Santiago}, \& {Youakim}}]{Carrillo2019}
{Carrillo}, I., {Minchev}, I., {Steinmetz}, M., {et~al.} 2019, arXiv e-prints
  [\eprint[arXiv]{1903.01493}]

\bibitem[{{Chakrabarty}(2007)}]{Chakrabarty2007}
{Chakrabarty}, D. 2007, \aap, 467, 145

\bibitem[{{Clarke} {et~al.}(2019){Clarke}, {Wegg}, {Gerhard}, {Smith}, {Lucas},
  \& {Wylie}}]{Clarke2019}
{Clarke}, J.~P., {Wegg}, C., {Gerhard}, O., {et~al.} 2019, arXiv e-prints
  [\eprint[arXiv]{1903.02003}]

\bibitem[{{Debattista} {et~al.}(2002){Debattista}, {Gerhard}, \&
  {Sevenster}}]{Debattista2002}
{Debattista}, V.~P., {Gerhard}, O., \& {Sevenster}, M.~N. 2002, \mnras, 334,
  355

\bibitem[{{Dehnen}(1998)}]{Dehnen1998}
{Dehnen}, W. 1998, \aj, 115, 2384

\bibitem[{{Dehnen}(1999{\natexlab{a}})}]{Dehnen1999}
{Dehnen}, W. 1999{\natexlab{a}}, \aj, 118, 1190

\bibitem[{{Dehnen}(1999{\natexlab{b}})}]{Dehnen1999bar}
{Dehnen}, W. 1999{\natexlab{b}}, \apjl, 524, L35

\bibitem[{{Dehnen}(2000)}]{Dehnen2000}
{Dehnen}, W. 2000, \aj, 119, 800

\bibitem[{{Dehnen} \& {Binney}(1998)}]{DehnenBinney19982}
{Dehnen}, W. \& {Binney}, J.~J. 1998, \mnras, 298, 387

\bibitem[{{Englmaier} \& {Gerhard}(1999)}]{Englmaier1999}
{Englmaier}, P. \& {Gerhard}, O. 1999, \mnras, 304, 512

\bibitem[{{Famaey} {et~al.}(2005){Famaey}, {Jorissen}, {Luri}, {Mayor}, {Udry},
  {Dejonghe}, \& {Turon}}]{Famaey2005}
{Famaey}, B., {Jorissen}, A., {Luri}, X., {et~al.} 2005, \aap, 430, 165

\bibitem[{{Fragkoudi} {et~al.}(2019){Fragkoudi}, {Katz}, {White}, {Di Matteo},
  {Trick}, {Sormani}, {Khoperskov}, {Haywood}, {Hall{\'e}}, \&
  {G{\'o}mez}}]{Fragkoudi2019}
{Fragkoudi}, F., {Katz}, D., {White}, S.~D.~M., {et~al.} 2019, arXiv e-prints
  [\eprint[arXiv]{1901.07568}]

\bibitem[{{Friske} \& {Sch{\"o}nrich}(2019)}]{FriskeSchonrich2019}
{Friske}, J. \& {Sch{\"o}nrich}, R. 2019, arXiv e-prints
  [\eprint[arXiv]{1902.09569}]

\bibitem[{{Fux}(1999)}]{Fux1999}
{Fux}, R. 1999, \aap, 345, 787

\bibitem[{{Fux}(2001)}]{Fux2001}
{Fux}, R. 2001, \aap, 373, 511

\bibitem[{{Gaia Collaboration} {et~al.}(2018){Gaia Collaboration}, {Brown},
  {Vallenari}, {Prusti}, {de Bruijne}, {Babusiaux}, {Bailer-Jones}, {Biermann},
  {Evans}, {Eyer}, \& et~al.}]{Gaia}
{Gaia Collaboration}, {Brown}, A.~G.~A., {Vallenari}, A., {et~al.} 2018, \aap,
  616, A1

\bibitem[{{G{\'o}mez} {et~al.}(2012){G{\'o}mez}, {Minchev}, {Villalobos},
  {O'Shea}, \& {Williams}}]{Gomez2012}
{G{\'o}mez}, F.~A., {Minchev}, I., {Villalobos}, {\'A}., {O'Shea}, B.~W., \&
  {Williams}, M.~E.~K. 2012, \mnras, 419, 2163

\bibitem[{{Hunt} \& {Bovy}(2018)}]{Hunt2018}
{Hunt}, J.~A.~S. \& {Bovy}, J. 2018, \mnras, 477, 3945

\bibitem[{{Katz} {et~al.}(2018){Katz}, {Antoja}, {Romero-G{\'o}mez}, {Drimmel},
  {Reyl{\'e}}, {Seabroke}, {Soubiran}, {Babusiaux}, {Di Matteo}, \&
  et~al.}]{Katz2018}
{Katz}, D., {Antoja}, T., {Romero-G{\'o}mez}, M., {et~al.} 2018, \aap, 616, A11

\bibitem[{{Khanna} {et~al.}(2019){Khanna}, {Sharma}, {Tepper-Garcia}, {Bland
  -Hawthorn}, {Hayden}, {Asplund}, {Buder}, {Chen}, {De Silva}, {Freeman},
  {Kos}, {Lin}, {Martell}, {Simpson}, {Stello}, {Ting}, {Zucker}, \&
  {Zwitter}}]{Khanna}
{Khanna}, S., {Sharma}, S., {Tepper-Garcia}, T., {et~al.} 2019, arXiv e-prints,
  arXiv:1902.10113

\bibitem[{{Khoperskov} {et~al.}(2019){Khoperskov}, {Di Matteo}, {Gerhard},
  {Katz}, {Haywood}, {Combes}, {Berczik}, \& {Gomez}}]{Khoperskov2018}
{Khoperskov}, S., {Di Matteo}, P., {Gerhard}, O., {et~al.} 2019, \aap, 622, L6

\bibitem[{{Laporte} {et~al.}(2019){Laporte}, {Minchev}, {Johnston}, \&
  {G{\'o}mez}}]{Laporte2018}
{Laporte}, C.~F.~P., {Minchev}, I., {Johnston}, K.~V., \& {G{\'o}mez}, F.~A.
  2019, \mnras, 485, 3134

\bibitem[{{Li} \& {Shen}(2019)}]{Li2019}
{Li}, Z.-Y. \& {Shen}, J. 2019, arXiv e-prints [\eprint[arXiv]{1904.03314}]

\bibitem[{{Long} {et~al.}(2013){Long}, {Mao}, {Shen}, \& {Wang}}]{Long2013}
{Long}, R.~J., {Mao}, S., {Shen}, J., \& {Wang}, Y. 2013, \mnras, 428, 3478

\bibitem[{{McMillan}(2013)}]{McMillan2013}
{McMillan}, P.~J. 2013, \mnras, 430, 3276

\bibitem[{{Minchev} {et~al.}(2010){Minchev}, {Boily}, {Siebert}, \&
  {Bienayme}}]{Minchev2010}
{Minchev}, I., {Boily}, C., {Siebert}, A., \& {Bienayme}, O. 2010, \mnras, 407,
  2122

\bibitem[{{Minchev} {et~al.}(2007){Minchev}, {Nordhaus}, \&
  {Quillen}}]{Minchev2007}
{Minchev}, I., {Nordhaus}, J., \& {Quillen}, A.~C. 2007, \apjl, 664, L31

\bibitem[{{Minchev} {et~al.}(2009){Minchev}, {Quillen}, {Williams}, {Freeman},
  {Nordhaus}, {Siebert}, \& {Bienaym{\'e}}}]{Minchev2009}
{Minchev}, I., {Quillen}, A.~C., {Williams}, M., {et~al.} 2009, \mnras, 396,
  L56

\bibitem[{{Monari} {et~al.}(2017{\natexlab{a}}){Monari}, {Famaey}, {Fouvry}, \&
  {Binney}}]{Monari2017b}
{Monari}, G., {Famaey}, B., {Fouvry}, J.-B., \& {Binney}, J.
  2017{\natexlab{a}}, \mnras, 471, 4314 (M17)

\bibitem[{{Monari} {et~al.}(2018){Monari}, {Famaey}, {Minchev}, {Antoja},
  {Bienaym{\'e}}, {Gibson}, {Grebel}, {Kordopatis}, {McMillan}, {Navarro},
  {Parker}, {Quillen}, {Reid}, {Seabroke}, {Siebert}, {Steinmetz}, {Wyse}, \&
  {Zwitter}}]{Monari2018}
{Monari}, G., {Famaey}, B., {Minchev}, I., {et~al.} 2018, Research Notes of the
  American Astronomical Society, 2, 32

\bibitem[{{Monari} {et~al.}(2016){Monari}, {Famaey}, \& {Siebert}}]{Monari2016}
{Monari}, G., {Famaey}, B., \& {Siebert}, A. 2016, \mnras, 457, 2569

\bibitem[{{Monari} {et~al.}(2017{\natexlab{b}}){Monari}, {Famaey}, {Siebert},
  {Duchateau}, {Lorscheider}, \& {Bienaym{\'e}}}]{Monari2017a}
{Monari}, G., {Famaey}, B., {Siebert}, A., {et~al.} 2017{\natexlab{b}}, \mnras,
  465, 1443

\bibitem[{{Monari} {et~al.}(2017{\natexlab{c}}){Monari}, {Kawata}, {Hunt}, \&
  {Famaey}}]{Monari2017}
{Monari}, G., {Kawata}, D., {Hunt}, J.~A.~S., \& {Famaey}, B.
  2017{\natexlab{c}}, \mnras, 466, L113

\bibitem[{{P{\'e}rez-Villegas} {et~al.}(2017){P{\'e}rez-Villegas}, {Portail},
  {Wegg}, \& {Gerhard}}]{PerezVillegas2017}
{P{\'e}rez-Villegas}, A., {Portail}, M., {Wegg}, C., \& {Gerhard}, O. 2017,
  \apjl, 840, L2

\bibitem[{{Pomp{\'e}ia} {et~al.}(2011){Pomp{\'e}ia}, {Masseron}, {Famaey}, {van
  Eck}, {Jorissen}, {Minchev}, {Siebert}, {Sneden}, {L{\'e}pine}, {Siopis},
  {Gentile}, {Dermine}, {Pasquato}, {van Winckel}, {Waelkens}, {Raskin},
  {Prins}, {Pessemier}, {Hensberge}, {Fr{\'e}mat}, {Dumortier}, \&
  {Bienaym{\'e}}}]{Pompeia2011}
{Pomp{\'e}ia}, L., {Masseron}, T., {Famaey}, B., {et~al.} 2011, \mnras, 415,
  1138

\bibitem[{{Portail} {et~al.}(2017){Portail}, {Gerhard}, {Wegg}, \&
  {Ness}}]{Portail2017}
{Portail}, M., {Gerhard}, O., {Wegg}, C., \& {Ness}, M. 2017, \mnras, 465, 1621
  (P17)

\bibitem[{{Quillen} {et~al.}(2018{\natexlab{a}}){Quillen}, {Carrillo},
  {Anders}, {McMillan}, {Hilmi}, {Monari}, {Minchev}, {Chiappini}, {Khalatyan},
  \& {Steinmetz}}]{Quillen2018b}
{Quillen}, A.~C., {Carrillo}, I., {Anders}, F., {et~al.} 2018{\natexlab{a}},
  \mnras, 480, 3132

\bibitem[{{Quillen} {et~al.}(2018{\natexlab{b}}){Quillen}, {De Silva},
  {Sharma}, {Hayden}, {Freeman}, {Bland-Hawthorn}, {{\v Z}erjal}, {Asplund},
  {Buder}, {D'Orazi}, {Duong}, {Kos}, {Lin}, {Lind}, {Martell}, {Schlesinger},
  {Simpson}, {Zucker}, {Zwitter}, {Anguiano}, {Carollo}, {Casagrande}, {Cotar},
  {Cottrell}, {Ireland}, {Kafle}, {Horner}, {Lewis}, {Nataf}, {Ting}, {Watson},
  {Wittenmyer}, \& {Wyse}}]{Quillen2018a}
{Quillen}, A.~C., {De Silva}, G., {Sharma}, S., {et~al.} 2018{\natexlab{b}},
  \mnras, 478, 228

\bibitem[{{Quillen} {et~al.}(2011){Quillen}, {Dougherty}, {Bagley}, {Minchev},
  \& {Comparetta}}]{Quillen2011}
{Quillen}, A.~C., {Dougherty}, J., {Bagley}, M.~B., {Minchev}, I., \&
  {Comparetta}, J. 2011, \mnras, 417, 762

\bibitem[{{Quillen} \& {Minchev}(2005)}]{Quillen2005}
{Quillen}, A.~C. \& {Minchev}, I. 2005, \aj, 130, 576

\bibitem[{{Ramos} {et~al.}(2018){Ramos}, {Antoja}, \& {Figueras}}]{Ramos2018}
{Ramos}, P., {Antoja}, T., \& {Figueras}, F. 2018, \aap, 619, A72

\bibitem[{{Rodriguez-Fernandez} \& {Combes}(2008)}]{Rodriguez2008}
{Rodriguez-Fernandez}, N.~J. \& {Combes}, F. 2008, \aap, 489, 115

\bibitem[{{Sanders} \& {Binney}(2016)}]{SandersBinney2016}
{Sanders}, J.~L. \& {Binney}, J. 2016, \mnras, 457, 2107

\bibitem[{{Sanders} {et~al.}(2019){Sanders}, {Smith}, \& {Evans}}]{Sanders2019}
{Sanders}, J.~L., {Smith}, L., \& {Evans}, N.~W. 2019, arXiv e-prints,
  arXiv:1903.02009

\bibitem[{{Sch{\"o}nrich} {et~al.}(2010){Sch{\"o}nrich}, {Binney}, \&
  {Dehnen}}]{Schonrich2010}
{Sch{\"o}nrich}, R., {Binney}, J., \& {Dehnen}, W. 2010, \mnras, 403, 1829

\bibitem[{{Sch{\"o}nrich} \& {Dehnen}(2018)}]{SchonrichDehnen2018}
{Sch{\"o}nrich}, R. \& {Dehnen}, W. 2018, \mnras, 478, 3809

\bibitem[{{Siebert} {et~al.}(2012){Siebert}, {Famaey}, {Binney}, {Burnett},
  {Faure}, {Minchev}, {Williams}, {Bienaym{\'e}}, {Bland-Hawthorn}, {Boeche},
  {Gibson}, {Grebel}, {Helmi}, {Just}, {Munari}, {Navarro}, {Parker}, {Reid},
  {Seabroke}, {Siviero}, {Steinmetz}, \& {Zwitter}}]{Siebert2012}
{Siebert}, A., {Famaey}, B., {Binney}, J., {et~al.} 2012, \mnras, 425, 2335

\bibitem[{{Sormani} {et~al.}(2015){Sormani}, {Binney}, \&
  {Magorrian}}]{Sormani2015}
{Sormani}, M.~C., {Binney}, J., \& {Magorrian}, J. 2015, \mnras, 454, 1818

\bibitem[{{Tremaine} \& {Weinberg}(1984)}]{Tremaine1984}
{Tremaine}, S. \& {Weinberg}, M.~D. 1984, \apj, 282, L5

\bibitem[{{Trick} {et~al.}(2019){Trick}, {Coronado}, \& {Rix}}]{Trick2018}
{Trick}, W.~H., {Coronado}, J., \& {Rix}, H.-W. 2019, \mnras, 484, 3291

\bibitem[{{Vauterin} \& {Dejonghe}(1997)}]{VauterinDejonghe1997}
{Vauterin}, P. \& {Dejonghe}, H. 1997, \mnras, 286, 812

\bibitem[{{Wegg} {et~al.}(2015){Wegg}, {Gerhard}, \& {Portail}}]{Wegg2015}
{Wegg}, C., {Gerhard}, O., \& {Portail}, M. 2015, \mnras, 450, 4050

\bibitem[{{Weinberg}(1994)}]{Weinberg1994}
{Weinberg}, M.~D. 1994, \apj, 420, 597

\bibitem[{{Weiner} \& {Sellwood}(1999)}]{Weiner1999}
{Weiner}, B.~J. \& {Sellwood}, J.~A. 1999, \apj, 524, 112

\end{thebibliography}

\end{document}